\documentclass{aa}

\usepackage{graphicx}
\usepackage{txfonts}

\usepackage{multirow}

\usepackage{natbib}
\usepackage[colorlinks=true,linkcolor=blue,citecolor=blue]{hyperref}
\bibpunct{(}{)}{;}{a}{}{,} 
	
\title{Effects of galaxy environment on merger fraction\thanks{Table \ref{tab:group} is only available in electronic form at the CDS via anonymous ftp to cdsarc.u-strasbg.fr (130.79.128.5) or via http://cdsweb.u-strasbg.fr/cgi-bin/qcat?J/A+A/}}

\author{W.~J.~Pearson\inst{\ref{inst:NCBJ}}
		\and D.~J.~D.~Santos\inst{\ref{inst:MPE}}
		\and T.~Goto\inst{\ref{inst:NTHU}}
		\and T.-C.~Huang\inst{\ref{inst:DSAS}, \ref{inst:ISAS}}
		\and S.~J.~Kim\inst{\ref{inst:NTHU}}
		\and H.~Matsuhara\inst{\ref{inst:DSAS}, \ref{inst:ISAS}}
		\and A.~Pollo\inst{\ref{inst:NCBJ}}
		\and S.~C.-C.~Ho\inst{\ref{inst:RSAA}, \ref{inst:CAS}, \ref{inst:OzGrav}, \ref{inst:ASTRO3D}}
		\and H.~S.~Hwang\inst{\ref{inst:AP-SNU}, \ref{inst:SNU}}
		\and K.~Małek\inst{\ref{inst:NCBJ}}
		\and T.~Nakagawa\inst{\ref{inst:ISAS}}
		\and M.~Romano\inst{\ref{inst:NCBJ}, \ref{inst:INAF}}
		\and S.~Serjeant\inst{\ref{inst:OU}}
		\and L.~Suelves\inst{\ref{inst:NCBJ}}
		\and H.~Shim\inst{\ref{inst:KNU}}
		\and G.~J.~White\inst{\ref{inst:RAL}, \ref{inst:OU}}
}

\institute{National Centre for Nuclear Research, Pasteura 7, 02-093 Warszawa, Poland\label{inst:NCBJ}\\\email{william.pearson@ncbj.gov.pl}
	\and Max Planck Institute for Extraterrestrial Physics, Gießenbachstraße 1, D-85748 Garching, Germany\label{inst:MPE}
	\and Institute of Astronomy, National Tsing Hua University, 101, Section 2. Kuang-Fu Road, Hsinchu, 30013, Taiwan\label{inst:NTHU}
	\and Department of Space and Astronautical Science, Graduate University for Advanced Studies, SOKENDAI, Shonankokusaimura, Hayama, Miura District, Kanagawa 240-0193, Japan\label{inst:DSAS}
	\and Institute of Space and Astronautical Science, Japan Aerospace Exploration Agency, 3-1-1 Yoshinodai, Chuo-ku, Sagamihara, Kanagawa 252-5210, Japan\label{inst:ISAS}
	\and Research School of Astronomy and Astrophysics, The Australian National University, Canberra, ACT 2611, Australia\label{inst:RSAA}
	\and Centre for Astrophysics and Supercomputing, Swinburne University of Technology, P.O. Box 218, Hawthorn, VIC 3122, Australia\label{inst:CAS}
	\and OzGrav: The Australian Research Council Centre of Excellence for Gravitational Wave Discovery, Hawthorn, VIC 3122, Australia\label{inst:OzGrav}
	\and ASTRO3D: The Australian Research Council Centre of Excellence for All-sky Astrophysics in 3D, ACT 2611, Australia\label{inst:ASTRO3D}
	\and Astronomy Program, Department of Physics and Astronomy, Seoul National University, 1 Gwanak-ro, Gwanak-gu, Seoul 08826, Republic of Korea\label{inst:AP-SNU}
	\and SNU Astronomy Research Center, Seoul National University, 1 Gwanak-ro, Gwanak-gu, Seoul 08826, Republic of Korea\label{inst:SNU}
	\and INAF - Osservatorio Astronomico di Padova, Vicolo dell’Osservatorio 5, I-35122, Padova, Italy\label{inst:INAF}
	\and School of Physical Sciences,The Open University, Walton Hall, Milton Keynes, MK7 6AA, England, UK\label{inst:OU}
	\and Department of Earth Science Education, Kyungpook National University, 80 Daehak-ro, Buk-gu, Daegu 41566, Republic of Korea\label{inst:KNU}
	\and Space Science and Technology Division, RALSpace, STFC Rutherford Appleton Laboratory, Chilton, Didcot, OX11 0QX, England, UK\label{inst:RAL}
}

\date{Received DD Month YYYY; accepted DD Month YYYY}

\abstract{}
{In this work, we intend to examine how environment influences the merger fraction, from the low density field environment to higher density groups and clusters. We also aim to study how the properties of a group or cluster, as well as the position of a galaxy in the group or cluster, influences the merger fraction.}
{We identified galaxy groups and clusters in the North Ecliptic Pole using a friends-of-friends algorithm and the local density. Once identified, we determined the central galaxies, group radii, velocity dispersions, and group masses of these groups and clusters. Merging systems were identified with a neural network as well as visually. With these, we examined how the merger fraction changes as the local density changes for all galaxies as well as how the merger fraction changes as the properties of the groups or clusters change.}
{We find that the merger fraction increases as local density increases and decreases as the velocity dispersion increases, as is often found in literature. A decrease in merger fraction as the group mass increases is also found. We also find groups with larger radii have higher merger fractions. The number of galaxies in a group does not influence the merger fraction.}
{The decrease in merger fraction as group mass increases is a result of the link between group mass and velocity dispersion. Hence, this decrease of merger fraction with increasing mass is a result of the decrease of merger fraction with velocity dispersion. The increasing relation between group radii and merger fraction may be a result of larger groups having smaller velocity dispersion at a larger distance from the centre or larger groups hosting smaller, infalling groups with more mergers. However, we do not find evidence of smaller groups having higher merger fractions.}

\keywords{Galaxies: clusters: general -- Galaxies: groups: general -- Galaxies: interactions -- Galaxies: statistics -- Galaxies: evolution}

\begin{document}
\authorrunning{W.~J.~Pearson et al.}
\maketitle

\section{Introduction}\label{sec:intro}

Galaxy groups are gravitationally bound structures that typically contain a few to a few tens of galaxies. These structures typically have halo masses between 10$^{11}$ and 10$^{14}$~M$_{\odot}$ \citep[e.g.]{2007ApJ...655..790C, 2017MNRAS.470.2982L, 2019MNRAS.490.2367C}. More massive structures with a few tens to a few hundreds of galaxies are considered to be galaxy clusters. These structures are understood to arise from the merger of smaller dark matter halos, or the capture of a smaller dark matter halo into a larger halo, in the current dark matter cosmology. Groups are thought to process galaxies before they form the larger clusters and super-clusters \citep[e.g.][]{2013MNRAS.432..336W, 2018MNRAS.474..547K}. Approximately half of all galaxies are believed to lie within a group or cluster, with the remaining galaxies found in the less dense environments as lone, field galaxies \citep[e.g.][]{2004MNRAS.348..866E, 2017MNRAS.470.2982L}.

The environment in which a galaxy lies is known to influence the physical properties of the galaxy. The typical star-formation rate of a galaxy is seen to be reduced for galaxies in denser environments when compared to the typical star-formation rate of a field galaxy of a similar stellar mass. This reduction in star-formation rate increases with the local density (i.e. the number of galaxies within a fixed radius) of a galaxy, with galaxies in more massive groups and clusters having lower star-formation rates than galaxies in less massive groups. Proximity to the centre of a group or cluster also influences star-formation rate, with lower star-formation rates seen for galaxies closer to the centre of a group or cluster \citep[e.g.][]{2002MNRAS.334..673L, 2003ApJ...584..210G, 2004MNRAS.353..713K, 2005ApJ...629..143B, 2006PASP..118..517B, 2009ARA&A..47..159B, 2009ApJ...699.1595P, 2013MNRAS.428.3306W, 2016ApJ...825..113D, 2018MNRAS.479.2147M, 2022A&A...666A..84G}. This environmental quenching is believed to be a result of interactions between the group or cluster members removing gas from the galaxies, as well as ram-pressure stripping as a galaxy moves through the group or cluster or starvation where the cluster dynamics prevent a galaxy accreting more gas to replenish exhausted reserves \citep[e.g.][]{1972ApJ...176....1G, 1980ApJ...237..692L, 1996Natur.379..613M, 1999MNRAS.308..947A, 2009ApJ...699.1595P, 2013MNRAS.432..336W, 2015Natur.521..192P, 2017MNRAS.469.3670S}. This removal of gas from the galaxies also results in the star-formation occurring closer to the centre of a galaxy in higher mass groups and clusters \citep[e.g.][]{2017MNRAS.464..121S, 2019MNRAS.483.2851S, 2022A&ARv..30....3B}. It has also been found that the hydrodynamical interactions (fluid interactions of the inter stellar medium and not gravitational interactions) between elliptical and spiral galaxies in clusters is a main driver of star-formation quenching in spiral galaxies \citep{2009ApJ...691.1828P, 2009ApJ...699.1595P, 2018ApJ...856..160H}. This is also seen for elliptical-spiral interactions in general \citep{2011A&A...535A..60H, 2016ApJS..222...16C, 2022ApJS..261...34H}.

The environment also influences the morphology of a galaxy. Groups and clusters are seen to have a higher fraction of elliptical galaxies than the field, which is seen to have a higher fraction of spiral galaxies. The exact split between the fraction of elliptical and spiral galaxies in groups or clusters and the field is uncertain. However, it is typical to find at least half of group galaxies to be elliptical compared to less than two-fifths of field galaxies \citep[e.g.][]{2007ApJ...670..190H, 2012ApJ...746..160W, 2013A&A...555A...5N, 2018MNRAS.481.3456C, 2020ApJ...898...20C}. Work by \citet{2021A&A...646A.151P} has found that galaxies become larger in more dense environments, a trend that is stronger for the largest members of a group. However, this same study did not find a change to the asymmetry of a galaxy when comparing field and group galaxies. As galaxies undergoing a merger can see an increase in asymmetry \citep[e.g.][]{2003AJ....126.1183C, 2022A&A...661A..52P, 2024MNRAS.527.6506B}, this would imply that there are not more, or fewer, mergers in group environments compared to the field.

Galaxy mergers are also a natural result of hierarchical growth found in the current cold dark matter cosmology. As the dark matter halos merge so will the baryonic counterparts \citep[e.g.][]{2014ARA&A..52..291C, 2015ARA&A..53...51S}. Much like being in different environments, galaxy mergers also act to change the morphologies and physical processes in the merging objects. The tidal forces between the two, or more, merging galaxies results in material being moved around and between the galaxies. This disruption can move stars, dust, and gas from the outer disk to the central bulge \citep[e.g.][]{1972ApJ...178..623T, 2015ARA&A..53...51S}.

This motion of material can result in the formation of shocks in the gas which can cause periods of intense star-formation: a starburst. These energetic events are believed to be the driving force behind some of the brightest infrared objects: ultra luminous infrared galaxies \citep[e.g.][]{1985MNRAS.214...87J, 1996ARA&A..34..749S, 2015A&A...577A.119C}. This has led to the idea that all major merging galaxies will go through a period of highly enhanced star-formation. However, the high star-formation rates are believed to be short-lived and are only observed in the minority of galaxies. Recent observations have found that galaxy mergers typically see star-formation rates increased by a factor of approximately two when compared to similar non-merging galaxies \citep[e.g.][]{2008AJ....135.1877E, 2011A&A...535A..60H, 2012MNRAS.426..549S, 2013MNRAS.435.3627E, 2013MNRAS.433L..59P, 2015MNRAS.454.1742K, 2019A&A...631A..51P, 2022ApJ...940....4S, 2022MNRAS.516.4922R}.

Star-formation is linked to the metallicity and stellar mass of a galaxy through the fundamental metallicity relation \citep[FMR,][]{2010MNRAS.408.2115M}. In this relation, galaxies with higher star-formation rates have lower metallicity at fixed stellar mass, thus galaxy mergers may be expected to have lower metallicity than non-merging galaxies due to higher star-formation rates. While this is seen, the metallicities of merging galaxies are found to be lower than predicted by the FMR \citep{2015MNRAS.451.4005G}. This lower than predicted metallicity is understood to be a result of low metallicity gas from the outskirts of the merging galaxies being driven into the centre as the merger proceeds \citep{2010ApJ...710L.156R, 2018MNRAS.479.3381B, 2017MNRAS.467.3898C, 2022MNRAS.509.2720S}. Galaxy pairs are also seen to have lower metallicity the closer their proximity \citep[e.g.][]{2008AJ....135.1877E, 2012MNRAS.426..549S}. Merger triggered starbursts are also known to affect scaling relations in simulations \citep[e.g.][]{2009MNRAS.400.1347C, 2011MNRAS.413L...1C, 2010MNRAS.407..749G, 2022MNRAS.515.3555P}

The motion of material towards the centre of the merging galaxies could also trigger an increase in active galactic nuclei (AGN), although the connection between mergers and AGN activity is still debated \citep[e.g.][]{1985AJ.....90..708K, 2011ApJ...743....2S, 2012A&A...538A..15H, 2012ApJ...744..148K, 2016ApJ...830..156M, 2017MNRAS.464.3882W, 2019MNRAS.487.2491E, 2020A&A...637A..94G, 2021ApJ...909..124S, 2023MNRAS.519.6149B}. Flyby interactions, where two or more galaxies interact but do not merge, can also drive material to the centre of galaxies and trigger AGN activity and star-formation \citep{1996Natur.379..613M, 1998ApJ...495..139M, 2023ApJ...954L...2C, 2024A&A...683A..57R}.

For cluster galaxies, it was often cited that the velocity dispersion is too high to allow mergers. This is a result of the galaxies approaching with too high a relative velocity to coalesce. This is despite the higher density resulting in a greater incidence of interactions \citep{1980ComAp...8..177O, 2004cgpc.symp..277M, 2018ApJS..237...14O}. Clusters, therefore, are typically seen to have fewer mergers than the field environment \citep[e.g.][]{2012MNRAS.425.2313T, 2017ApJ...843..126D}. The outer regions of clusters, where the velocity dispersion is potentially lower, can have galaxies approaching slow enough to result in mergers \citep{2004cgpc.symp..277M, 2018ApJ...869....6D, 2019MNRAS.486..868K}. This may be either due to accreting slower field galaxies; accreting galaxy groups, which have lower velocity dispersions \citep{2020MNRAS.498.3852B}; or accreting ongoing mergers \citep{2018ApJS..237...14O}. Clusters can also accrete recently completed mergers, that still show morphological disturbances caused by the merger event \citep{2012ApJS..202....8S, 2013A&A...554A.122Y, 2018ApJS..237...14O}. There are also fewer galaxy pairs, often used as a way to identify galaxy mergers, found in higher density environments and clusters. This is a result of the relative velocity requirement used for galaxies pairs, $\Delta v < 500~\mathrm{km~s}^{-1}$, being lower than the typical relative velocities in clusters, $500 < \Delta v < 1000~\mathrm{km~s}^{-1}$ \citep{2010MNRAS.407.1514E}. However due to their higher density (number of galaxies within a fixed radius), groups and clusters are found to have higher merger fractions than field galaxies \citep{2017MNRAS.469.4551K, 2019MNRAS.486..868K}.

This work aims to study how galaxy environment influences the fraction of galaxies that are undergoing a merger in the 5.4 sq. deg. North Ecliptic Pole field (NEP). The NEP field has an extensive wealth of multi-wavelength data, from the ultra-violet (UV) to radio, including deep Hyper Suprime-Cam (HSC) imaging. The high quality optical imaging allows robust galaxy merger detection, while the extensive wavelength coverage allows accurate estimations of the galaxies' physical properties. 
This study compares how various environmental properties of galaxies influence the merger fraction: local density as well as the group or cluster occupation number, radius, velocity dispersion and mass. Groups and clusters will be found through a friends-of-friends algorithm. This will determine the number of galaxies in the group or cluster and allow the calculation of the other physical properties of these structures. This will allow us to examine how each of these properties' impact on the merger fraction are related to the other physical properties, a first such study. Identifying the central and satellite galaxies in the group or cluster will also allow us to examine if these different galaxies' merger fractions are influenced in different ways.

The paper is structured as follows. In Sect. \ref{sec:data}, the data used will be discussed along with merger identification and the detection of groups and clusters. Section \ref{sec:group} describes how the velocity dispersion, group or cluster radius and group or cluster mass are calculated. Results are presented in Sect. \ref{sec:results} and discussion is held in Sect. \ref{sec:discuss}. Finally, we conclude in Sect. \ref{sec:conc}. Where necessary, we use the Planck 2015 cosmology: $\Omega_{m}$ = 0.307, $\Omega_{\Lambda}$ = 0.693 and $H_{0}$ = 67.7 km~s$^{-1}$~Mpc$^{-1}$\citep{2016A&A...594A..13P}

\section{Data}\label{sec:data}
The merger fraction in different galaxy environments are derived within NEP, covering an area of 5.4 sq. deg., where the multi-wavelength (ultra violet to sub-mm) data \citep{2021MNRAS.500.4078K} are derived from follow-up observations based on the AKARI's NEP survey \citep{2012A&A...548A..29K}. Where necessary, photometry from the Hyper Suprime-Cam - NEP survey is used \citep[HSC-NEP;][]{2017PKAS...32..225G, 2018PASJ...70S...1M, 2021MNRAS.500.5024O}. Redshifts used in the work are those of \citet{2021MNRAS.506.6063H}. These are a combination of spectroscopic and photometric redshifts, with the photometric redshifts being used where the spectroscopic redshifts are not available. The photometric redshifts were derived using the Canada France Hawaii Telescope MegaPrime $u$-band \citep{2003SPIE.4841...72B, 2014A&A...566A..60O}, HSC $g$, $r$, $i$, $z$, and $Y$-bands \citep{2021MNRAS.502..140H}, and the Spitzer Infrared Array Camera bands 1 and 2 \citep{2004ApJS..154...10F, 2018ApJS..234...38N} using \texttt{LePhare} \citep{1999MNRAS.310..540A, 2006A&A...457..841I}. Further discussion of the photometric redshifts can be found in \citet{2021MNRAS.506.6063H}. Spectroscopic redshifts are obtained using optical spectroscopic follow up programs for mid-infrared sources \citep{2013ApJS..207...37S, 2021MNRAS.506.6063H}. Once we apply our redshift cut of $z \leq 0.3$, 736 of our 34\,264 galaxies (2.1\%) have a spectroscopic redshift.

\subsection{Merger identification}
For this work, we use the merging galaxies identified by \citet{2022A&A...661A..52P}. This catalogue is a hybrid deep-learning-human approach, using a deep neural network to identify merger candidates that were then subsequently visually inspected by astronomers to confirm, or deny, their status as a galaxy merger. This was done using HSC $r$-band imaging data.

The merger candidates were selected using a combined convolutional neural network, on the galaxy images, and a fully connected network, on the morphologies of the galaxies. The resulting network has an accuracy of 88.4\% at $z < 0.15$ and 85.0\% at $0.15 \leq z < 0.30$ and expects to recover 86.3\% and 79.0\% of mergers at $z < 0.15$ and $0.15 \leq z < 0.30$, respectively. All merger candidates were visually inspected for evidence of merger features. The human classifiers have an accuracy of approximately 62.5\% based on identifying galaxy mergers from simulations that had not already been pre-selected by the neural network. As a result of using non-preselected galaxies from simulations, the accuracy of the visual selection may not represent the true ability of the human classifiers to identify mergers in pre-selected observations. Further details of this merger selection can be found in \citet{2022A&A...661A..52P}.

Of the 34\,264 galaxies at $z < 0.3$ that form our sample, the catalogue provides 10\,195 major merger candidates (29.8\%) from the neural network classification of which 2109 (6.2\% of the full sample or 20.7\% of the merger candidates) were visually confirmed as galaxy mergers. The merger identification is limited to $z < 0.3$ due to the poor performance of the deep neural networks at higher redshifts. As a result, this work has the same redshift limit applied.

In this work, we examine both the merger candidates and the visually confirmed mergers separately. A neural network may correctly identify merging galaxies that a human would not, or cannot. The network also does not include redshift information, so can identify chance projections as mergers. Thus, while the visually confirmed merger sample should be more pure, the merger candidate sample should be more complete. As a result, we also use the merger candidates with the understanding that the sample will be more contaminated with non-mergers than the visually confirmed sample. We define merger fraction as the number of merging galaxies divided by the total number of galaxies (both mergers and non-mergers).

As galaxy mergers are expected to experience an increase in star-formation rate during the merger, there may also be an increase in radio emission from the galaxies. This radio emission arises due to massive stars (M $>8$~M$_{\odot}$) emitting ionising radiation producing free-free (thermal) radio emission in HII regions \citep[e.g.][]{1992ARA&A..30..575C}. When these short-lived, massive stars become supernovae, the supernovae remnants accelerate electrons and release synchrotron radiation \citep[e.g.][]{1992ARA&A..30..575C, 2004A&A...427..525B}. Cross matching sources detected within an area of 1.9 sq. deg. by the Giant Metrewave Radio Telescope \citep[GMRT,][]{2017PKAS...32..231W} and detected within an area of 10 sq. deg. by the Low-Frequency Array \citep[LOFAR,][]{2013A&A...556A...2V, WhiteLOFAR} to our sample of galaxies using a matching radius of 3'', we find 104 of the 1675 GMRT sources and 266 of the 14\,673 LOFAR sources to match with our catalogue. From these cross matches, we find 53 (51\%) and 157 (59\%) of the GMRT and LOFAR sources, respectively, to be merger candidates. This drops to 9 (9\%) and 31 (12\%) of the radio sources being visually confirmed mergers. A larger fraction of the radio sources are identified as merger candidates or visually confirmed mergers than our parent sample, suggesting that merging galaxies are more active at radio wavelengths than non-merging galaxies. Of the merger candidates, 0.5\% are detected by GMRT and 1.5\% are detected by LOFAR while for the visually confirmed mergers these percentages only drop slightly to 0.4\% and 1.4\%. These relatively consistent fractions may be due to the small number of cross matches. Alternatively, it may be the result of the deep neural network identifying galaxy mergers that humans find hard to identify, and are therefore rejected during visual classification.

The increase in star-formation rate expected in galaxy mergers can also result in an increase in infrared emission. To test this, we match our sample of galaxies with the 5.4 sq. deg. AKARI NEP band-matched catalogue \citep{2012A&A...548A..29K} within 3''. AKARI was an infrared space telescope \citep{2007PASJ...59S.369M}. Of the 114\,794 AKARI sources, 7259 match to our NEP galaxy catalogue (here after AKARI galaxies). From the AKARI galaxies, 2289 (31.5\%) are identified as merger candidates and 466 (6.4\%) are visually confirmed mergers. These fractions of merger candidates and visually confirmed mergers of AKARI galaxies are similar to the fractions of merger candidates and visually confirmed mergers found for our sample of galaxies being used in this work.  Of our sample of merger candidates and visually confirmed mergers, 22.4\% and 22.1\% have AKARI cross matches, respectively, similar to the fraction of galaxies in our sample that are matched to AKARI (21.2\%). These percentages suggest that the galaxy mergers do not preferentially produce an excess of infrared emission. If the merging galaxies preferentially had an excess of infrared emission, we would expect a larger fraction of the AKARI galaxies to be merging compared to our sample. Similarly, we would expect a larger percentage of our merger candidates and visually confirmed mergers to be detected by AKARI compared to the percentage of galaxies from the whole sample that are detected by AKARI if the mergers had an infrared excess.

\subsection{Normalised local density}
\begin{figure}
	\resizebox{\hsize}{!}{\includegraphics{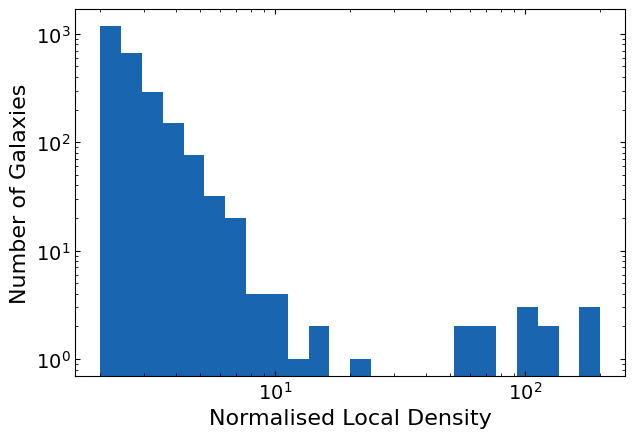}}
	\caption{Distribution of the number of galaxies that have a certain normalised local density.}
	\label{fig:dist:density}
\end{figure}

For the normalised local density, we use the results obtained by \citet{2021MNRAS.507.3070S}, also used in \citet{2021MNRAS.506.6063H}. In \citet{2021MNRAS.507.3070S} the local densities are derived using the 10$^{th}$ nearest neighbour in 2-dimensions using only galaxies within a redshift bin of width $0.065 \times (1+z)$. Edge corrections are applied for galaxies near the edge of the survey and the density is normalised to the median density within the redshift bin. Galaxies are determined to be in an over-density if the normalised density is greater than 2. Here we require that a group or cluster galaxy to be in an over-density \citep{2021MNRAS.506.6063H}. Full details of how the local density is determined can be found in \citet{2021MNRAS.507.3070S} while the distribution of local densities is presented in Fig. \ref{fig:dist:density}.

\subsection{Friends of friends}
To find groups and clusters, we apply a friends-of-friends algorithm to the galaxies in NEP that lie in an over-density, closely following \citet{2021MNRAS.506.6063H} with some changes to allow the identification of smaller structures. A friend of a galaxy is defined as a galaxy that lies within a linking length on the sky and has a redshift that satisfies \citep{2021MNRAS.506.6063H}
\begin{equation}
	|(z_{o}-z_{f})/(1+z_{o})| < 0.032,
\end{equation}
where $z_{o}$ is the redshift of the origin galaxy and $z_{f}$ is the redshift of the (potential) friend galaxy.

A variable linking length is used, as suggested by \citet{2014A&A...566A...1T}, which takes the form
\begin{equation}
	f(z) = d_{0} \times \big(1 + \alpha_{0} \mathrm{arctan} \bigg(\frac{z}{z_{+}} \bigg) \big),
\end{equation}
where we take the values of $d_{0}$, $\alpha_{0}$ and $z_{+}$ from \citet{2021MNRAS.506.6063H} as 0.146, 0.867 and 0.088, respectively. As \citet{2021MNRAS.506.6063H} use the same galaxies, redshifts and local densities as this work, these values are deemed suitable.

Where we deviate from \citet{2021MNRAS.506.6063H} is in identifying the groups and clusters. To find groups and clusters, all galaxies that have at least four other galaxies as friends are found, compared to nine in \citet{2021MNRAS.506.6063H}. This allows us to identify smaller structures with fewer galaxies. We chose five as a threshold as it is roughly half way between two (a pair) and ten (as used in \citet{2021MNRAS.506.6063H}). These friend-of-friend groups are then merged where they contain common galaxies to create groups and clusters. As a friend-of-friend group must have at least five members, all groups will contain at least five galaxies. Further distinction between groups and clusters is not made. Thus, for easier communication, both groups and clusters will be referred to as groups with the understanding that this label may be inaccurate. Galaxies that are not in a group are considered to be field galaxies.

\begin{figure}
	\resizebox{\hsize}{!}{\includegraphics{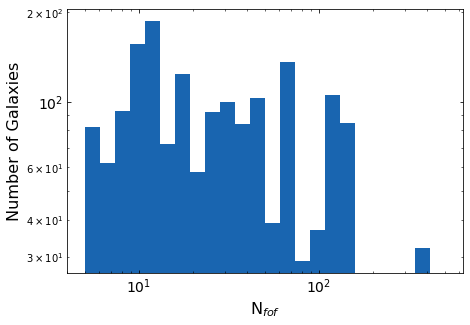}}
	\caption{Distribution of the number of galaxies at $z < 0.3$ that lie in a group with N$_{fof}$ members.}
	\label{fig:dist:n-fof}
\end{figure}

To prevent group galaxies being missed at the edge of our redshift range ($z = 0.3$), and hence the potential for groups to be missed or split, the friends-of-friends algorithm is applied to all galaxies out to $z = 1.1$. This results in 1767 galaxies identified as part of 188 groups at $z < 0.3$. We note that not all members of these 188 groups have $z < 0.3$ and so the total number of galaxies in these 188 groups is larger than 1767. Figure \ref{fig:dist:n-fof} shows the distribution of the number of galaxies at $z < 0.3$ within groups with a certain number of friends of friends (N$_{fof}$) members. All the group properties are calculated using the entire group, not just the group members at $z < 0.3$.

\subsection{Mass completeness}
For this study, we require a mass complete sample. To determine the mass completeness, we follow \citet{2022A&A...661A..52P} and perform the empirical mass completeness estimate for galaxies at $z < 0.3$ as determined using \citep{2010A&A...523A..13P}:
\begin{equation}
\log(M_{lim}) = \log(M_{\star}) - 0.4(r_{lim} - r),
\end{equation}
where $M_{\star}$ is the observed stellar mass of the galaxy in M$_{\odot}$, $r_{lim}$ is the limiting $r$-band magnitude, here 26, $r$ is the measured $r$-band magnitude, and $M_{lim}$ is the lowest mass that can be observed for the galaxy at $r_{lim}$. The limiting mass is then the value for $M_{lim}$ that 90\% of the faintest 20\% have masses below. The observed masses of the galaxies were determined using spectral energy density fitting with \texttt{LePhare} \citep{1999MNRAS.310..540A, 2006A&A...457..841I, 2021MNRAS.502..140H}. Using this process, the limiting mass of our galaxies was determined to be $10^{8.04}$~M$_{\odot}$, leaving 22\,310 galaxies for further study.

\section{Group properties}\label{sec:group}
The group properties were calculated, and are presented here in Table \ref{tab:group}, before the mass completeness cut was applied.

\begin{table*}
	\centering
	\caption{Properties of the group a galaxy is a member of.}
	\label{tab:group}
	\begin{tabular}{ccccccc}
		\hline
		\hline
		HSC\_ID & Group & N$_{fof}$ & Central & R$_{50}$ & $\sigma$ & M$_{50}$ \\
		& & & & Mpc & kms$^{-1}$ & log(M$_{\odot}$) \\
		\hline
		79671321218284477 & N/A & N/A & N/A & N/A & N/A & N/A \\
		80093795676346180 & N/A & N/A & N/A & N/A & N/A & N/A \\
		80092971042649458 & N/A & N/A & N/A & N/A & N/A & N/A \\
		79666652588832016 & 176 & 7 & False & 65.72 & 313.25 & 16.18 \\
		79671329808217055 & 1 & 16 & False & 157.13 & 524.25 & 17.0 \\
		79217648117756584 & 30 & 8 & False & 46.28 & 1171.89 & 17.17 \\
		79666240271977762 & 85 & 19 & False & 41.33 & 961.46 & 16.95 \\
		79671737830102246 & 102 & 7 & False & 83.66 & 1496.84 & 17.64 \\
		79675989847731138 & 138 & 7 & False & 11.72 & 48.42 & 13.81 \\
		79675989847731369 & 138 & 7 & False & 11.72 & 48.42 & 13.81 \\
		... & ... & ... & ... & ... & ... & ... \\
		\hline
	\end{tabular}
	\tablefoot{If value is N/A, galaxy is a field galaxy.\\The full table is available at the CDS.}
\end{table*}

\subsection{Central galaxy and group radius}
The central galaxy of a group is determined iteratively following \citet{2011MNRAS.416.2640R}. The HSC-NEP $r$-band centre of light of the group is determined in 3-dimensional space. The galaxy that is furthest from the centre of light is removed and the position of the centre of light is recalculated. This process is repeated until only two galaxies are left. The central galaxy is then chosen to be the brightest of the two galaxies in the $r$-band. The remaining galaxies in the group are then taken to be satellite galaxies. We find 114 central galaxies at $z < 0.3$, and hence 1653 satellite galaxies in our redshift range before we apply the mass completeness limit. Of our mass complete sample, 1192 are group galaxies of which 103 are found to be central galaxies and 1089 are satellites.

\begin{figure}
	\resizebox{\hsize}{!}{\includegraphics{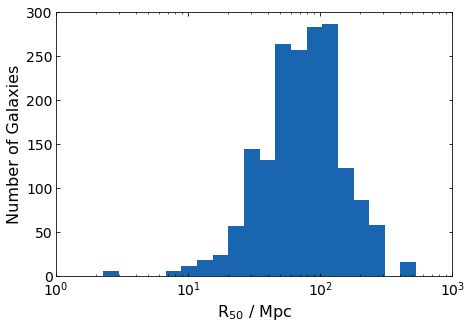}}
	\caption{Distribution of the number of galaxies at $z < 0.3$ that lie in a group with a certain R$_{50}$.}
	\label{fig:dist:r-50}
\end{figure}

Once the central galaxy has been identified, the radius of the group is then the distance from the central galaxy within which 50\% of the galaxies in the group lie, which shall hence forth be referred to as R$_{50}$. This radius was computed in 3-dimensional space, with the component perpendicular to the image plane calculated as the difference between the angular diameter distance of the central galaxy to the other galaxies in the group. The distribution of the number of galaxies that lie in a group with a certain R$_{50}$ can be found in Fig. \ref{fig:dist:r-50}

\subsection{Velocity Dispersion}
\begin{figure}
	\resizebox{\hsize}{!}{\includegraphics{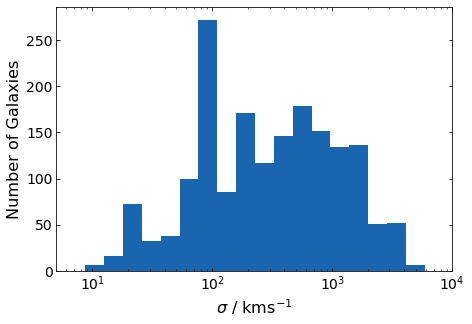}}
	\caption{Distribution of the number of galaxies at $z < 0.3$ that lie in a group with a certain velocity dispersion.}
	\label{fig:dist:sigma}
\end{figure}

The velocity dispersion of the groups is determined using the gapper estimator \citep{1990AJ....100...32B}. The recession velocity ($v_{rec}$) of all galaxies within a group, with $N$ members, is determined from the redshift of the galaxy. These $v_{rec}$ are then ordered and the relative velocities between each pair ($g$) is calculated in order along with their corresponding weights ($w$). That is $g_{i} = v_{rec, i+1} - v_{rec, i}$ and $w_{i} = i(N-i)$ for $i = 1, 2, ..., N-1$. The gapper velocity dispersion ($\sigma_{gap}$) is then calculated as
\begin{equation}
	\sigma_{gap} = \frac{\sqrt{\pi}}{N(N-1)} \sum_{i=1}^{N} g_{i}w_{i}.
\end{equation}
As the brightest galaxy is expected to move with the centre of mass of the host halo, the velocity dispersion will increase by an extra factor of $\sqrt{\frac{N}{N-1}}$ \citep{2004MNRAS.348..866E, 2011MNRAS.416.2640R}. As the velocity dispersion will be influenced by uncertainty, which we take to be 10\% of $\sigma_{gap}$, we subtract the expected error in quadrature \citep{2011MNRAS.416.2640R}. Thus, the velocity dispersion ($\sigma$) is calculated as
\begin{equation}
 \sigma = \sqrt{\frac{N}{N-1} - 0.01}~\sigma_{gap}.
\end{equation}
The distribution of galaxies in a group of a certain $\sigma$ can be found in Fig. \ref{fig:dist:sigma}.

\subsection{Group mass}
\begin{figure}
	\resizebox{\hsize}{!}{\includegraphics{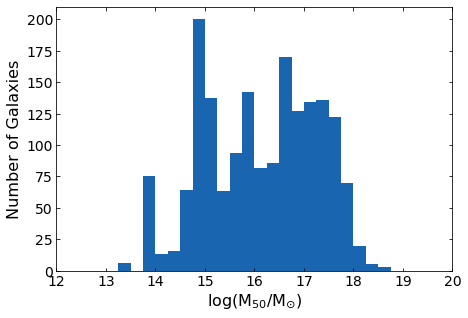}}
	\caption{Distribution of the number of galaxies at $z < 0.3$ that lie in groups of different masses (M$_{50}$).}
	\label{fig:dist:mass}
\end{figure}

To determine the mass of the groups, we assume that the groups are in virial equilibrium. Thus the mass of the group can be estimated with
\begin{equation}
	M_{50} = \frac{A}{G} \sigma^{2} R_{50},
\end{equation}
where M$_{50}$ is in M$_{\odot}$, $A$ is a scaling factor required to create a median-unbiased mass estimate, $G$ is the gravitational constant, $\sigma$ is in km~s$^{-1}$ and R$_{50}$ is in Mpc. We take the value of $A$ to be 10 \citep{2011MNRAS.416.2640R}. Figure \ref{fig:dist:mass} shows the distribution of galaxies that lie in groups of different M$_{50}$.

As can be seen in Fig. \ref{fig:dist:mass}, we have a large number of galaxies that lie in very high mass groups (log(M$_{50}$/M$_{\odot}$) > 16). These high group masses are a result of large group radii found in Fig. \ref{fig:dist:r-50}. As we primarily use photometric redshifts in our friend-of-friends algorithm, it is likely that the large R$_{50}$, and hence large M$_{50}$, are a result of the lower precision of photometric redshifts, compared to spectroscopic redshifts. This lower precision may result in galaxies being assigned to groups where they do not truly belong. Our friend-of-friends algorithm had a false detection rate of $\approx1\%$ when finding clusters with at least ten members and has been shown to have a linking length that is not unreasonably large \citep{2021MNRAS.506.6063H}. However, \citet{2021MNRAS.506.6063H} do caution that using the friend-of-friends algorithm to find fewer than 10 friends may result in over-assignment to a group, which may also be causing our large structures. We choose to keep the massive groups for further analysis.

\section{Results}\label{sec:results}
\subsection{Correlation between group properties}
As we would expect the properties that are correlated to have similar trends for how the merger fraction evolves, we examine which properties are correlated. For this we determine the Pearson coefficient (Pc) when comparing the properties in linear-linear space, log-linear space, linear-log space and log-log space. If the magnitude of Pc is less than 0.1, we determine there is no correlation, between 0.1 and 0.5 to be weakly correlated (or anti-correlated), between 0.5 and 0.9 to be correlated, and between 0.9 and 1.0 to be strongly correlated. We report the strongest (anti-)correlations, along with their p-values.

We find that the normalised local density does not correlate with any other property. For N$_{fof}$, we find it to be correlated with R$_{50}$ in log-log space ($\mathrm{Pc} = 0.512, \mathrm{p} = 0.000$), anti-correlated with $\sigma$ in log-log space ($\mathrm{Pc} = -0.634, \mathrm{p} = 0.000$), and weakly anti-correlated with M$_{50}$ in log-linear space (log(N$_{fof}$) and M$_{50}$, $\mathrm{Pc} = -0.344, \mathrm{p} = 0.000$). R$_{50}$ is weakly anti-correlated with $\sigma$ in log-linear space (log(R$_{50}$) and $\sigma$, $\mathrm{Pc} = -0.338, \mathrm{p} = 0.000$) while finally velocity dispersion is strongly correlated with M$_{50}$ in log-log space ($\mathrm{Pc} = 0.966, \mathrm{p} = 0.000$). Full correlations can be found in Appendix \ref{app:correlations}.

\subsection{Change in merger fraction}
In the following, we present our results for how the merger fraction changes as different group properties change. We perform the analysis after binning the galaxies in bins of width 0.4 for normalised local density, 20~Mpc for R$_{50}$, 200~km~s$^{-1}$ for the velocity dispersion and 1.0~log(M$_{\mathbf{\star}}$/M$_{\odot}$) for group mass. For N$_{fof}$, we bin in log-space with bins of width 0.1. These bin sizes were chosen to provide reasonable statistics. The trends are derived by fitting a straight line to the merger fractions in the bins which contain at least 10 galaxies. We require at least 10 galaxies in a bin to allow an accurate determination of the merger fraction. The uncertainties on the merger fractions in each bin are estimated with bootstrapping. For bins where the merger fraction is zero, the average error of the other bins is used as the uncertainty, as bootstrapping will always return an uncertainty of zero. 
The fitted parameters can be found in Appendix \ref{app:params}.

\subsubsection{Normalised local density}
\begin{figure}
	\resizebox{\hsize}{!}{\includegraphics{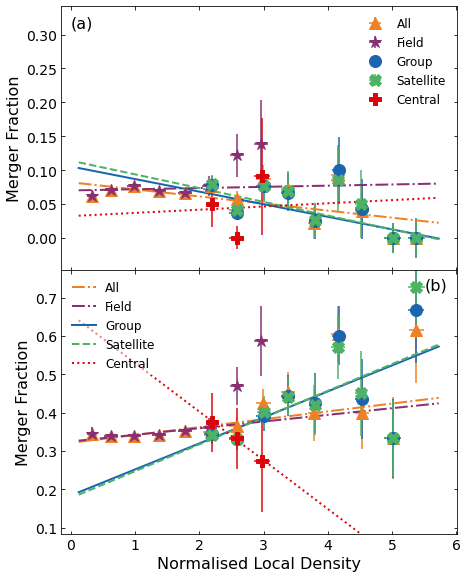}}
	\caption{Merger fraction as a function of normalised local density for (a) visually confirmed mergers and (b) merger candidates. The entire galaxy sample (orange triangles), field galaxies (purple stars), group galaxies (blue circles), satellite galaxies (green crosses) and central galaxies (red pluses) are shown, along with their fitted linear trends. Best fit parameters can be found in Appendix \ref{app:params}.}
	\label{fig:mgr:density}
\end{figure}

In Fig. \ref{fig:mgr:density} we show the trend of merger fraction as a function of normalised local density. For the deep-learning merger candidates, we see an increase in merger fraction with normalised local density (at $>$~3 standard deviations (SD)) while the visually selected mergers see a slight decrease in merger fraction as the environment becomes denser for all galaxies (at $>$~3SD). For field galaxies, there is a very slight increase in the merger fraction as the normalised local density increases for both the merger candidates ($>$~1SD) and the visually selected mergers ($>$~3SD). As the entire population is seen to have a reduction in visually selected merger fraction, it is not surprising to see a reduction in the visually selected merger fraction for the group galaxies ($>$~2SD). Similarly, a strong increase in merger fraction is seen for the merger candidates for group galaxies ($>$~2SD).

Splitting the group galaxies further, into central and satellite galaxies, we find a decrease in visually selected merger fraction with normalised local density for the satellite galaxies ($>$~2SD) while the central galaxies are consistent with no change. For the merger candidates, we find that the central galaxies have a decrease in merger fraction as the normalised local density increases ($>$~8SD) while the satellites have an increasing merger fraction with normalised local density ($>$~3SD).

\subsubsection{Friends of friends}
\begin{figure}
	\resizebox{\hsize}{!}{\includegraphics{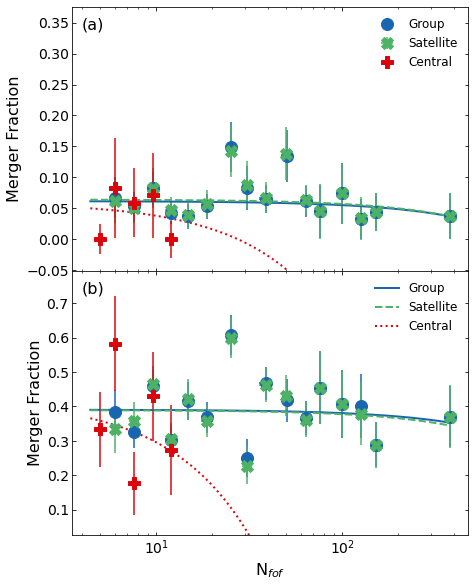}}
	\caption{Merger fraction as a function of N$_{fof}$ for (a) visually confirmed mergers and (b) merger candidates. The group galaxies (blue circles), satellite galaxies (green crosses) and central galaxies (red pluses) are shown, along with their fitted linear trends. The trend fitting was performed with a linear trend ($y = mx + c$) to log-binned data. Best fit parameters can be found in Appendix \ref{app:params}.}
	\label{fig:mgr:n-fof}
\end{figure}

For the group galaxies, we find that the merger fraction is consistent with being constant with N$_{fof}$, or equivalently the number of galaxies in a group. This relation is seen for the satellite galaxies and the central galaxies, for both the merger candidates and the visually confirmed mergers as seen in Fig. \ref{fig:mgr:n-fof}.

\subsubsection{Group radius}
\begin{figure}
	\resizebox{\hsize}{!}{\includegraphics{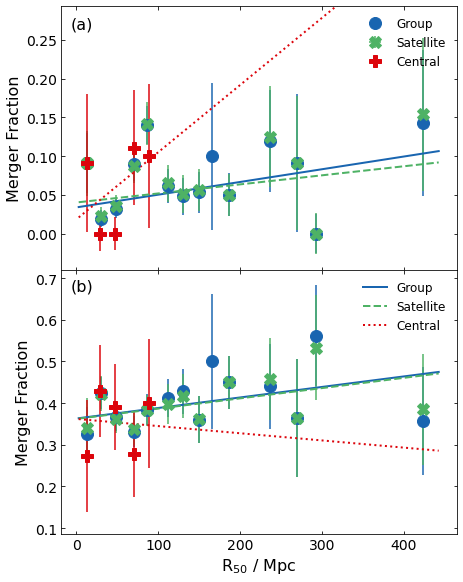}}
	\caption{Merger fraction as a function of group radius for (a) visually confirmed mergers and (b) merger candidates. The group galaxies (blue circles), satellite galaxies (green crosses) and central galaxies (red pluses) are shown, along with their fitted linear trends. Best fit parameters can be found in Appendix \ref{app:params}.}
	\label{fig:mgr:r-50}
\end{figure}

The radius of a group is seen to influence the merger fraction, as shown in Fig. \ref{fig:mgr:r-50}. The larger R$_{50}$ is, the higher the fraction of merging galaxies for both the merger candidates ($>$~1SD) and the visually confirmed mergers ($>$~1SD). For the satellite galaxies, the fraction of merger candidates slightly increases with R$_{50}$ ($>$~1SD). The fraction of visually selected mergers, again for the satellite galaxies, also shows a very slight increase with R$_{50}$ ($<$~1SD) but is consistent with no change. The fraction of merger candidates decreases with increasing R$_{50}$ ($<$~1SD) while the visually selected merger fraction is consistent with no change.

\subsubsection{Velocity dispersion}
\begin{figure}
	\resizebox{\hsize}{!}{\includegraphics{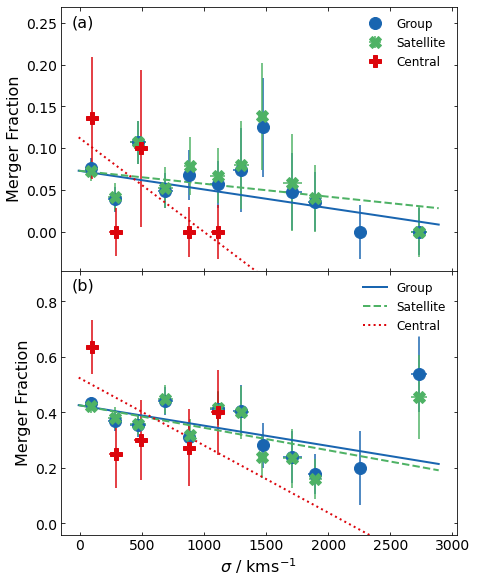}}
	\caption{Merger fraction as a function of velocity dispersion for (a) visually confirmed mergers and (b) merger candidates. The group galaxies (blue circles), satellite galaxies (green crosses) and central galaxies (red pluses) are shown, along with their fitted linear trends. Best fit parameters can be found in Appendix \ref{app:params}.}
	\label{fig:mgr:velocity}
\end{figure}

The merger fraction of the group galaxies in Fig. \ref{fig:mgr:velocity} is seen to decrease as the velocity dispersion increases for the merger candidates ($>$~2SD) and visually confirmed mergers ($>$~2SD). For the satellite galaxies, the merger candidates and visually confirmed mergers both see a decrease in merger fraction as the velocity dispersion increases ($>$~2SD and $>$~1SD respectively). The central galaxies again suffer from a small number of bins with at least 10 galaxies within them. Once fitted, we find that the merger fraction decreases as $\sigma$ increases for the merger candidates ($>$~1SD) and the visually confirmed mergers ($>$~1SD).

\subsubsection{Group mass}
\begin{figure}
	\resizebox{\hsize}{!}{\includegraphics{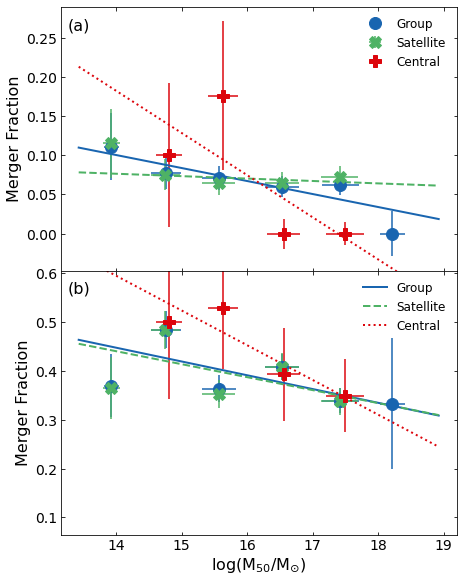}}
	\caption{Merger fraction as a function of group mass for (a) visually confirmed mergers and (b) merger candidates. The group galaxies (blue circles), satellite galaxies (green crosses) and central galaxies (red pluses) are shown, along with their fitted linear trends. Best fit parameters can be found in Appendix \ref{app:params}.}
	\label{fig:mgr:mass}
\end{figure}

For the final environmental factor we consider, the group mass, we find that the visually confirmed mergers and the merger candidates have a decreasing merger fraction as M$_{50}$ increases ($>$~2SD and $>$~1SD respectively), as seen in Fig. \ref{fig:mgr:mass}. The visually confirmed central galaxies ($>$~1SD) and merger candidate centrals ($>$~3SD) and satellites ($>$~1SD) follow the same trend while the visually confirmed satellite galaxies are consistent with having no change in merger fraction with M$_{50}$.

\section{Discussion}\label{sec:discuss}
\subsection{Normalised local density}
It is generally seen and understood that galaxy mergers are more common in denser environment if velocity dispersion is low enough to allow galaxies to become gravitationally bound to one-another. This is what is seen in this work for the merger candidates, with the merger fraction increasing for the entire galaxy population as the density increases, but not for the visually confirmed mergers. This is likely a result of the merger candidates being contaminated by galaxies that are close on the sky but otherwise unassociated in higher density environments. These unassociated galaxies are then removed in the visually confirmed sample. As the entire galaxy population contains group galaxies, which are higher density environments and are expected to have fewer mergers due to the high velocity dispersion, it makes sense that the entire population will see an overall decrease in merger fraction as the density increases.

Indeed, removing the group galaxies and only examining the field galaxies does agree with this idea: both the merger candidates and the visually confirmed mergers show an increase of merger fraction with normalised local density for the field galaxies. Similarly, the visually confirmed mergers show a decrease in merger fraction in groups, as would be expected. On the other hand, the merger candidates see an increase in merger fraction for group galaxies as the normalised local density increases, again suggesting contamination in the sample. As the visually selected mergers see a decrease in merger fraction for group galaxies, it is not surprising to find the satellite galaxies having a decrease in merger fraction as their normalised local density increases. With the central galaxies showing no change in merger fraction with the normalised local density, it is therefore the satellite galaxies that are driving the trend for the entire group galaxy population.

For the merger candidates, the increase in merger fraction with density for field galaxies is again driven by the satellite galaxies, with the central galaxies seeing a decrease in merger fraction for higher density environments. This would suggest that, for the merger candidates, the expected increase in velocity dispersion with normalised local density is only influencing the central galaxies while it only affects the satellite galaxies for the visually confirmed mergers.

As noted above, an increase in merger fraction with the local density is seen in other works, although direct quantitative comparisons are difficult due to different definitions of local density. We compare our work to a few examples here. Galaxies from HSC Subaru Strategic Program \citep[HSC-SSP][]{2022PASJ...74..247A} have been identified as merging systems using Zoobot, a machine learning galaxy classifier \citep{2023JOSS....8.5312W}, at redshifts less than 0.3 \citep{2023A&A...679A.142O}. Using a local density defined as the distance to the fifth nearest neighbour, these galaxies showed an increase in merger fraction as density increased. This study does not, however, consider separation between field and group galaxies or central and satellite group galaxies.

A similar trend has also been observed at higher redshift. Using the pair fraction, that is the fraction of galaxies that are observed to be close in projection and have relative velocities less than 500~kms$^{-1}$. \citet{2010ApJ...718.1158L} found the number of close pairs increases with local density, here defined to be the third nearest neighbour, at redshifts of $0.75 < z < 1.2$ in the DEEP2 Redshift Survey \citep{2003SPIE.4834..161D}. Converting the pair fraction to a merger rate, the same study finds an increasing merger rate with local density, despite the fraction of close pairs expected to merge decreasing with local density. The merger rate can be calculated from the merger fraction using a merger time. Therefore, to first order if a constant merger time is assumed, the merger rate can be considered to be a scaled merger fraction. \citet{2010ApJ...718.1158L} note that due to the redshift completeness, the third nearest neighbour local density observed can be considered to be the sixth nearest neighbour local density. Results from semi analytic models implimented in the Millennium Simulation \citep{2005Natur.435..629S} have shown that the pair fraction and merger rate increase with normalised local density \citep{2012ApJ...754...26J}. In \citet{2012ApJ...754...26J} the local density is defined for the sixth nearest neighbour.

Not all studies find an increasing merger fraction with local density. A citizen science project GALAXY CRUISE \citep{2023PASJ...75..986T} has found a decrease in the fraction of interacting galaxies as the local density decreases for galaxies at $z < 0.1$. Local density is defined in \citet{2023PASJ...75..986T} for the fifth and twentieth nearest neighbours. \citet{2023PASJ...75..986T} conclude that their trend is not dominated by groups due to their trend holding for local densities derived for both fifth and twentieth nearest neighbours. We note that \citet{2023PASJ...75..986T} do not separate field from group galaxies in their study. This agrees with our findings of the merger fraction for all galaxies changing with local density independently of the merger fraction in group galaxies.

For mergers in groups, it has been shown observationally that mergers are more likely to take place in less dense groups \citep{2012A&A...539A..46A}. This agrees with our findings for the visually confirmed mergers but not the merger candidates. This further supports the idea that the merger candidates' contamination by non-merging galaxies is exacerbated in more dense groups.

\subsection{Friends of friends}
The number of galaxies in a group does not influence the number of galaxies that are merging. Thus, any changes in the merger fraction are not due to the size of a group but must instead be due to the other properties of a group, such as density, $\sigma$ or M$_{50}$. It may be expected that more misclassified mergers in the candidate sample would be found in groups with a higher number of galaxies due to chance projections. This could result in a positive correlation between the merger candidate fraction and N$_{fof}$. Evidently, this is not the case. N$_{fof}$ has a weak anti-correlation with the group mass (log(N$_{fof}$) and M$_{50}$, Pc = -0.344, p = 0.000) and correlated with group radius (log(N$_{fof}$) and 
log(R$_{50}$), Pc = 0.212, p = 0.000) so it may be expected that these two properties have a similar lack of merger fraction change for group and satellite galaxies. Merger fraction as a function of the number of galaxies in a group in not well studied in previous works.

\subsection{Group radius}
The radius of the group influences the merger fraction of galaxies. A larger R$_{50}$, and hence a physically larger group, will contain a larger fraction of merging galaxies, as can be seen in Fig. \ref{fig:mgr:r-50} where both samples are characterized by a similar increasing trend with R$_{50}$. This may be due to there being more galaxies in the outer regions of the group, where the velocity dispersion is believed to be lower and thus the galaxies are more easily able to merge \citep{2004cgpc.symp..277M, 2018ApJ...869....6D, 2019MNRAS.486..868K}. If smaller groups have a larger $\sigma$ gradient with radius, there are fewer pairs of galaxies that are at similar enough velocities to be able to merge. Alternatively, as it has been found that smaller groups falling into larger groups, or clusters, contain the majority of merging galaxies \citep{2020MNRAS.498.3852B}, a physically larger group is more likely to have an infalling group containing the merging systems. As a result, groups with larger R$_{50}$ could have a higher fraction of merging systems. If this was true, we would also expect the smallest groups to also have higher merger fractions, which we do not see in this work. The central galaxies are the only subsample to show a decrease in merger fraction as R$_{50}$ increases, for the merger candidates, possibly due to the expected increase in $\sigma$ in the centre of larger groups or an increase in chance projections for smaller groups increasing the merger fraction at lower R$_{50}$. It is also possible that the relation is flat for merger candidates, as the uncertainties are too significant to fit this relation precisely.

R$_{50}$ is weakly anti-correlated with $\sigma$ (log(R$_{50}$) and $\sigma$, Pc = -0.338, p = 0.000) so the increase in merger fraction as R$_{50}$ increases may be a result of a lower $\sigma$ in larger galaxy groups. Similarly, R$_{50}$ is correlated with N$_{fof}$, with larger R$_{50}$ potentially resulting in a larger number of galaxies in the group. With more galaxies available to merge in physically larger groups, it may be that the increase in merger fraction with R$_{50}$ could be simply a result of more galaxies in the group. With more galaxies available to merge, there is a greater chance that two galaxies have low enough relative velocity to merge and hence have a higher merger fraction. However, if this were true we would likely see an increase in merger fraction as N$_{fof}$ increases that, as discussed above, we do not see. Merger fraction as a function of group radius in a group in not well studied in previous works.

\subsection{Velocity dispersion}
As $\sigma$ increases, it is generally seen that the merger fraction decreases, as expected \citep[e.g.][]{1980ComAp...8..177O, 2004cgpc.symp..277M, 2021gfe..book..181C}. This would be a result of the high relative velocities of the galaxies resulting in interactions and fly-bys but not mergers. This trend continues for both the central and satellite galaxies for the merger candidates and the visually confirmed mergers. The outer regions of a group are expected to have a lower velocity dispersion than the inner regions \citep{2004cgpc.symp..277M, 2018ApJ...869....6D, 2019MNRAS.486..868K}. This difference is seen as the trend of decreasing merger fraction is less extreme with the satellite galaxies than the central galaxies, suggesting the outer regions have a lower $\sigma$ and are more easily able to host mergers. Velocity dispersion in the groups is highly correlated with group mass, as would be expected as M$_{50}$ is derived from $\sigma$, but not R$_{50}$. Thus, we would expect to see similar results with M$_{50}$.

A study with the European Southern Observatory Distant Cluster Survey \citep{2005A&A...444..365W} found limited changes to the merger fraction with $\sigma$. \citet{2018ApJ...869....6D} identified merging and tidally interacting galaxies using a Gini-M$_{20}$ cut \citep{2004AJ....128..163L, 2008ApJ...672..177L} tuned with visual classification. Using the spectroscopically identified groups, \citet{2018ApJ...869....6D} found only weak evidence for a decrease in merger fraction as $\sigma$ increases. When \citet{2018ApJ...869....6D} also included the photometrically identified groups, they found no change in the merger fraction as $\sigma$ increased. In this work, we use primarily photometrically identified groups, suggesting that we should not find a change in merger fraction with $\sigma$. However, our merger sample is selected differently, with mergers identified with deep-learning and then visually confirmed. Thus it is likely that our merger selection method is causing the differences in trends found, suggesting that our method finds fewer galaxy mergers in high $\sigma$ environments. Why this may be is unclear but potentially could be that the Gini-M$_{20}$ cut incorrectly selects galaxies that are interacting, but not merging, in the high $\sigma$ groups.

\subsection{Group mass}
The visually confirmed merger fraction and the merger candidate merger fraction are seen to decrease as M$_{50}$ increases, for all group galaxies and the central galaxies. For the satellites, however, we see no change in merger fraction with the group mass of the for the visually confirmed mergers but we do see a decrease for the merger candidates. It is not clear where this difference arises from for the satellite galaxies. One may expect higher mass groups to have more galaxies and, as a result, have more chance projections. If this were the case, we would expect the merger candidates to have no change in merger fraction and the visually confirmed mergers to have a decrease in the merger fraction as M$_{50}$ increases. Evidently, this is not what is seen. Alternatively, if smaller groups are denser, there is a greater probability of chance projections but, again, we do not see lower mass groups having a higher density environment. Generally, however, the trends for the majority of the samples are similar to what is seen for $\sigma$ suggesting that the trends for all but the visually confirmed mergers are driven by $\sigma$ and its link to M$_{50}$.

Data from the Sloan Digital Sky Survey \citep[SDSS,][]{2004AJ....128..502A} has been used to trace the merger rate as a function of group virial mass between 10$^{13.4}$ and 10$^{14.9}$~M$_{\odot}$ for massive galaxies (merging pairs with a combined mass greater than 10$^{11}$~M$_{\odot}$) at $z < 0.12$. \citet{2008MNRAS.388.1537M} saw that the merger fraction remained approximately constant with group mass along with the fraction of central galaxies undergoing a merger. Using close pairs, instead of morphologically disturbed galaxies, the same study found an increase in merger fraction for both the central galaxies and all group galaxies with group mass. Both of these results are not what we see, with decreases in merger fraction for central and all group galaxies. However, we cover a larger range of group masses, which may be the cause of the differences.

A different observational study using the Zurich Environmental Study database \citep{2013ApJ...776...71C}, derived from the Percolation Inferred Galaxy Group catalogue of the Two-degree-Field Galaxy Redshift Survey \citep{2001MNRAS.328.1039C, 2004MNRAS.348..866E}, did find a change in the merger fraction as the group mass changes. With galaxies at $0.05 < z < 0.0585$, \citet{2014ApJ...797..127P} found lower mass groups to contain a higher fraction of merging galaxies compared to high mass groups, by a factor of $\approx2.5$. This factor is larger than seen in this work, possibly a result of our study using a larger range of redshifts. The use of spectroscopic redshifts by \citet{2014ApJ...797..127P}, compared to our primarily photometric redshifts, may also be a cause of this larger factor. The merger selection method used in \citet{2014ApJ...797..127P} differs from this work. \citet{2014ApJ...797..127P} identified mergers as morphologically disturbed or irregular single galaxies, close pairs, or close and morphologically disturbed pairs. In comparison, this work used deep-learning merger candidates or those candidates which were visually confirmed. As a result, we may be missing some close pairs that are not morphologically disturbed in lower mass groups or \citet{2014ApJ...797..127P} may be selecting close pairs that are not merging in higher mass groups.

Simulations also show change in the merger fraction with halo mass. Semi analytic models implemented in the Millennium simulation found an increase in the merger rate at lower halo masses ($<10^{13.7}$~M$_{\odot}$), followed by a decrease at higher halo masses ($>10^{13.7}$~M$_{\odot}$) \citep{2012ApJ...754...26J}. This agrees with what is seen in our work. Our group masses are typically large ($>10^{13}$~M$_{\odot}$) and we see a reduction in merger fraction as the mass increases.

\section{Conclusions}\label{sec:conc}
In this work, we study how the environment a galaxy lies in and the physical properties of a group influence the merger fraction of galaxies in the North Ecliptic Pole. For this, we study the normalised local density, for group and field galaxies, and the number of galaxies in a group, the group radius, the velocity dispersion of the group, and mass of the group and observe how the merger fraction changes as these properties change. This is a first of its kind study to investigate these properties together. The groups are determined using a friends-of-friends algorithm on galaxies in over-densities. We find the following:
\begin{enumerate}[i]
	\item The merger fraction increases as normalised local density increases: 0.02 merger candidate fraction increase per unit increase in normalised local density.
	\item The number of galaxies in a group does not influence the merger fraction.
	\item Groups with a larger radius have higher merger fractions: 1.64e-4 visually confirmed merger fraction increase per Mpc.
	\item The higher the velocity dispersion, the lower the merger fraction: 2.22e-5 visually confirmed merger fraction decrease per kms$^{-1}$.
	\item The higher the group mass, the lower the merger fraction: 1.66e-3 visually confirmed merger fraction decrease per log(M$_{50}$/M$_{\odot}$).
\end{enumerate}

Results i and iv were expected and often found in literature. We suggest that iii is a result of a slower change in $\sigma$ as a galaxy is further out in a group, allowing galaxies to be at similar enough velocities to merge. Alternatively, iii may be a result of larger groups being able to host smaller, infalling groups that are found to have more mergers, although we do not find a significant increase in merger fraction for smaller groups, both in radius and in galaxy number. Result v is a result of the group mass being derived from the velocity dispersion: the decreasing merger fraction with $\sigma$ is causing the decreasing merger fraction with M$_{50}$.

Future work with a larger sample of galaxies, for example with the Vera C. Rubin Observatory, will provide better statistics. The Vera C. Rubin Observatory will cover approximately 18\,000 sq. deg. of the sky, an area over 3000 time larger than NEP but with a comparable depth \citep{2017arXiv170804058L}. Thus, it may be expected to find over 56\,000 groups at the same redshifts as this work. With so many more groups, the changes in merger fraction with environmental properties will be better constrained and we will be able to study the change in merger fraction for central galaxies over a wider range of the parameters. The current number of galaxies is too low to study the redshift evolution of the trends we find. The larger Vera C. Rubin Observatory survey will provide orders of magnitude more galaxies. This will allow the redshift dependence of the environmental properties examined here to be explored, something that cannot be done in NEP due to the current, relatively small sample size. This will provide insight into if the environmental impact on merging galaxies had redshift dependence.




\begin{acknowledgements}
We would like to thank the referee for their thorough and thoughtful comments that helped improve the quality and clarity of this paper.
	
W.J.P. has been supported by the Polish National Science Center project UMO-2020/37/B/ST9/00466.

This research was conducted under the agreement on scientific cooperation between the Polish Academy of Sciences and the Ministry of Science and Technology in Taipei and supported by the Polish National Science Centre grant UMO-2018/30/M/ST9/00757 and by Polish Ministry of Science and Higher Education grant DIR/WK/2018/12.

SH acknowledges the support of The Australian Research Council Centre of Excellence for Gravitational Wave Discovery (OzGrav) and the Australian Research Council Centre of Excellence for All Sky Astrophysics in 3 Dimensions (ASTRO 3D), through project number CE17010000 and CE170100013, respectively.

HSH acknowledges the support by the National Research Foundation of Korea (NRF) grant funded by the Korea government (MSIT) (No. 2021R1A2C1094577).

K. M. has been supported by the Polish National Science Centre project UMO-2018/30/E/ST9/00082.

TN acknowledges the support by JSPS KAKENHI Grants Number 21H04496 and 23H05441.

M.R. acknowledges support from the Narodowe Centrum Nauki (UMO-2020/38/E/ST9/00077) and support from the Foundation for Polish Science (FNP) under the program START 063.2023.

H.S. acknowledges the support from the National Research Foundation of Korea grant No.2021R1A2C4002725 and No.2022R1A4A3031306, funded by the Korea government (MSIT).

GJW gratefully acknowledges receipt of an Emeritus Fellowship from The Leverhulme Trust
\end{acknowledgements}

\bibliographystyle{aa} 
\bibliography{merger-in-groups} 

\begin{appendix}
\section{Group property correlations}\label{app:correlations}
Here we present the full Pearson coefficients in Table \ref{tab:app:pearson} and p-values in Table \ref{tab:app:pvalue} between all parameters.

\begin{table*}
	\centering
	\caption{Pearson coefficients when comparing parameters in columns with parameters in rows in linear and log space.}
	\label{tab:app:pearson}
	\begin{tabular}{cc|cccccccccc}
		\hline
		\hline
		
		 & & \multicolumn{2}{c}{Normalised} & \multicolumn{2}{c}{\multirow{2}{*}{N$_{fof}$}} & \multicolumn{2}{c}{\multirow{2}{*}{R$_{50}$}} & \multicolumn{2}{c}{\multirow{2}{*}{$\sigma$}} & \multicolumn{2}{c}{\multirow{2}{*}{M$_{50}$}} \\
		 & & \multicolumn{2}{c}{local density} & & & & & & & & \\
		 & & Linear & Log & Linear & Log & Linear & Log & Linear & Log & Linear & Log \\
		\hline
		Normalised & Linear & - & - & -0.032 & -0.023 & -0.063 & -0.076 & -0.008 & 0.032 & -0.029 & 0.007 \\
		local density & Log & - & - & 0.015 & 0.077 & -0.022 & -0.009 & -0.070 & -0.051 & -0.056 & -0.003 \\
		
		\multirow{2}{*}{N$_{fof}$} & Linear & -0.032 & 0.015 & - & - & 0.317 & 0.332 & -0.359 & -0.498 & -0.212 & -0.344 \\
		 & Log & -0.023 & 0.077 & - & - & 0.445 & 0.512 & -0.564 & -0.634 & -0.101 & -0.118 \\
		 
		\multirow{2}{*}{R$_{50}$} & Linear & -0.063 & -0.022 & 0.317 & 0.445 & - & - & -0.262 & -0.236 & -0.048 & -0.030 \\
		 & Log & -0.076 & -0.009 & 0.332 & 0.512 & - & - & -0.338 & -0.297 & -0.011 & -0.029 \\
		 
		\multirow{2}{*}{$\sigma$} & Linear & -0.008 & -0.070 & -0.359 & -0.564 & -0.262 & -0.338 & - & - & 0.767 & 0.237 \\
		 & Log & 0.032 & -0.051 & -0.498 & -0.634 & -0.236 & -0.297 & - & - & 0.545 & 0.966 \\
		 
		\multirow{2}{*}{M$_{50}$} & Linear & -0.029 & -0.056 & -0.212 & -0.344 & -0.048 & -0.030 & 0.767 & 0.545 & - & - \\
		 & Log & 0.007 & -0.003 & -0.101 & -0.118 & -0.011 & -0.029 & 0.237 & 0.966 & - & - \\
		\hline
	\end{tabular}
\end{table*}

\begin{table*}
	\centering
	\caption{p-values when comparing parameters in columns with parameters in rows in linear and log space.}
	\label{tab:app:pvalue}
	\begin{tabular}{cc|cccccccccc}
		\hline
		\hline
		
		 & & \multicolumn{2}{c}{Normalised} & \multicolumn{2}{c}{\multirow{2}{*}{N$_{fof}$}} & \multicolumn{2}{c}{\multirow{2}{*}{R$_{50}$}} & \multicolumn{2}{c}{\multirow{2}{*}{$\sigma$}} & \multicolumn{2}{c}{\multirow{2}{*}{M$_{50}$}} \\
		 & & \multicolumn{2}{c}{local density} & & & & & & & & \\
		 & & Linear & Log & Linear & Log & Linear & Log & Linear & Log & Linear & Log \\
		\hline
		Normalised & Linear & - & - & 0.184 & 0.327 & 0.008 & 0.001 & 0.727 & 0.186 & 0.225 & 0.753 \\
		local density & Log & - & - & 0.529 & 0.001 & 0.348 & 0.703 & 0.003 & 0.032 & 0.019 & 0.913 \\
		
		\multirow{2}{*}{N$_{fof}$} & Linear & 0.184 & 0.529 & - & - & 0.000 & 0.000 & 0.000 & 0.000 & 0.000 & 0.000 \\
		 & Log & 0.327 & 0.001 & - & - & 0.000 & 0.000 & 0.000 & 0.000 & 0.000 & 0.000 \\
		
		\multirow{2}{*}{R$_{50}$} & Linear & 0.008 & 0.348 & 0.000 & 0.000 & - & - & 0.000 & 0.000 & 0.045 & 0.206 \\
		 & Log & 0.001 & 0.703 & 0.000 & 0.000 & - & - & 0.000 & 0.000 & 0.649 & 0.222 \\
		
		\multirow{2}{*}{$\sigma$} & Linear & 0.727 & 0.003 & 0.000 & 0.000 & 0.000 & 0.000 & - & - & 0.000 & 0.000 \\
		 & Log & 0.186 & 0.032 & 0.000 & 0.000 & 0.000 & 0.000 & - & - & 0.000 & 0.000 \\
		
		\multirow{2}{*}{M$_{50}$} & Linear & 0.225 & 0.019 & 0.000 & 0.000 & 0.045 & 0.206 & 0.000 & 0.000 & - & - \\
		 & Log & 0.753 & 0.913 & 0.000 & 0.000 & 0.649 & 0.222 & 0.000 & 0.000 & - & - \\
		\hline
	\end{tabular}
\end{table*}
	
\section{Fitted parameters}\label{app:params}
Here we present the parameters from the trend fitting in Tables \ref{tab:app:density}, \ref{tab:app:n-fof}, \ref{tab:app:r-50}, \ref{tab:app:velocity}, and \ref{tab:app:m-50}. The trends are fitted as a straight line of the form
\begin{equation}\label{eq:app:line}
	y = \alpha x + \beta
\end{equation}
where $y$ is the merger fraction, $x$ is the parameter being fitted, $\alpha$ is the slope, and $\beta$ is the normalisation. We fit using orthogonal distance regression \citep{boggs1990orthogonal, 2020SciPy-NMeth}.

\begin{table*}
	\centering
	\caption{Trend fitting parameters and their uncertainties for merger fraction as a function of normalised local density using Eq. \ref{eq:app:line}}
	\label{tab:app:density}
	\begin{tabular}{cc|cccc}
		\hline
		\hline
		Merger & Galaxy & $\alpha$ & $\sigma_{\alpha}$ & $\beta$ & $\sigma_{\beta}$ \\
		\hline
		\multirow{5}{*}{Visual} & All & -1.04e-02 & 2.88e-03 & 8.21e-02 & 4.28e-03\\
		 & Field & 1.77e-03 & 4.74e-03 & 7.01e-02 & 5.11e-03\\
		 & Group & -1.86e-02 & 8.04e-03 & 1.05e-01 & 2.55e-02\\
		 & Satellite & -2.03e-02 & 7.29e-03 & 1.14e-01 & 2.34e-02\\
		 & Central & 4.72e-03 & 1.00e-01 & 3.22e-02 & 2.38e-01\\
		\hline
		\multirow{5}{*}{Candidate} & All & 2.05e-02 & 6.58e-03 & 3.21e-01 & 8.27e-03\\
		 & Field & 1.74e-02 & 1.12e-02 & 3.25e-01 & 1.22e-02\\
		 & Group & 6.79e-02 & 2.32e-02 & 1.84e-01 & 6.62e-02\\
		 & Satellite & 6.99e-02 & 2.31e-02 & 1.77e-01 & 6.59e-02\\
		 & Central & -1.27e-01 & 1.49e-02 & 6.57e-01 & 3.70e-02\\
		\hline
	\end{tabular}
\end{table*}

\begin{table*}
	\centering
	\caption{Trend fitting parameters and their uncertainties for merger fraction as a function of N$_{fof}$ using Eq. \ref{eq:app:line}}
	\label{tab:app:n-fof}
	\begin{tabular}{cc|cccc}
		\hline
		\hline
		Merger & Galaxy & $\alpha$ & $\sigma_{\alpha}$ & $\beta$ & $\sigma_{\beta}$ \\
		\hline
		\multirow{3}{*}{Visual} & Group & -6.50e-05 & 8.73e-05 & 6.16e-02 & 7.81e-03\\
		 & Satellite & -7.14e-05 & 8.17e-05 & 6.41e-02 & 7.71e-03\\
		 & Central & -2.17e-03 & 8.02e-03 & 5.96e-02 & 6.78e-02\\
		\hline
		\multirow{3}{*}{Candidate} & Group & -9.95e-05 & 3.39e-04 & 3.91e-01 & 2.72e-02\\
		 & Satellite & -1.19e-04 & 3.47e-04 & 3.89e-01 & 2.83e-02\\
		 & Central & -1.25e-02 & 3.29e-02 & 4.21e-01 & 2.68e-01\\
		\hline
	\end{tabular}
\end{table*}

\begin{table*}
	\centering
	\caption{Trend fitting parameters and their uncertainties for merger fraction as a function of R$_{50}$ using Eq. \ref{eq:app:line}}
	\label{tab:app:r-50}
	\begin{tabular}{cc|cccc}
		\hline
		\hline
		Merger & Galaxy & $\alpha$ & $\sigma_{\alpha}$ & $\beta$ & $\sigma_{\beta}$ \\
		\hline
		\multirow{3}{*}{Visual} & Group & 1.64e-04 & 1.47e-04 & 3.37e-02 & 1.47e-02\\
		 & Satellite & 1.17e-04 & 1.45e-04 & 4.01e-02 & 1.55e-02\\
		 & Central & 8.67e-04 & 9.89e-04 & 1.83e-02 & 5.62e-02\\
		\hline
		\multirow{3}{*}{Candidate} & Group & 2.52e-04 & 1.70e-04 & 3.63e-01 & 1.96e-02\\
		 & Satellite & 2.46e-04 & 1.52e-04 & 3.62e-01 & 1.78e-02\\
		 & Central & -1.72e-04 & 1.57e-03 & 3.62e-01 & 8.53e-02\\
		\hline
	\end{tabular}
\end{table*}

\begin{table*}
	\centering
	\caption{Trend fitting parameters and their uncertainties for merger fraction as a function of velocity dispersion using Eq. \ref{eq:app:line}}
	\label{tab:app:velocity}
	\begin{tabular}{cc|cccc}
		\hline
		\hline
		Merger & Galaxy & $\alpha$ & $\sigma_{\alpha}$ & $\beta$ & $\sigma_{\beta}$ \\
		\hline
		\multirow{3}{*}{Visual} & Group & -2.22e-05 & 9.23e-06 & 7.29e-02 & 9.50e-03\\
		 & Satellite & -1.55e-05 & 1.12e-05 & 7.30e-02 & 9.79e-03\\
		 & Central & -1.12e-04 & 6.58e-05 & 1.12e-01 & 4.41e-02\\
		\hline
		\multirow{3}{*}{Candidate} & Group & -7.30e-05 & 2.99e-05 & 4.25e-01 & 2.44e-02\\
		 & Satellite & -8.10e-05 & 3.08e-05 & 4.26e-01 & 2.41e-02\\
		 & Central & -2.42e-04 & 2.20e-04 & 5.23e-01 & 1.31e-01\\
		\hline
	\end{tabular}
\end{table*}

\begin{table*}
	\centering
	\caption{Trend fitting parameters and their uncertainties for merger fraction as a function of log(M$_{50}$) using Eq. \ref{eq:app:line}}
	\label{tab:app:m-50}
	\begin{tabular}{cc|cccc}
		\hline
		\hline
		Merger & Galaxy & $\alpha$ & $\sigma_{\alpha}$ & $\beta$ & $\sigma_{\beta}$ \\
		\hline
		\multirow{3}{*}{Visual} & Group & -1.66e-02 & 6.63e-03 & 3.32e-01 & 1.09e-01\\
	 	 & Satellite & -3.07e-03 & 4.97e-03 & 1.19e-01 & 8.06e-02\\
		 & Central & -5.37e-02 & 3.50e-02 & 9.34e-01 & 5.64e-01\\
		\hline
		\multirow{3}{*}{Candidate} & Group & -2.82e-02 & 1.80e-02 & 8.43e-01 & 2.91e-01\\
		 & Satellite & -2.66e-02 & 2.37e-02 & 8.12e-01 & 3.82e-01\\
		 & Central & -7.12e-02 & 2.02e-02 & 1.59e+00 & 3.37e-01\\
		\hline
	\end{tabular}
\end{table*}

\end{appendix}

\end{document}